# A different explanation of energy-resolved scanning tunnelling results from $(Ca_{2-x}Na_x)CuO_2Cl_2$ than that suggested by Hanaguri *et al* (2009).

**John. A. Wilson**

H. H. Wills Physics Laboratory
University of Bristol
Bristol BS8 1TL.  U.K.

**Abstract**

The scanning tunnelling spectroscopy results from $(Ca_{2-x}Na_x)CuO_2Cl_2$ in a strong magnetic field are reinterpreted in a substantially different fashion.  Instead of looking, as Hanaguri *et al* do, to a $B_{1g}$ BCS-based interpretation and relying heavily upon 'coherence effects', the very detailed changes wrought in the tunnelling characteristics are re-addressed following the present author's negative-$U$, boson-fermion, resonant crossover modelling of the High Temperature Superconducting Cuprate (HTSC) phenomenon.  As with a great many other now quite sophisticated and discriminatory experimental assaults on the latter problem, it would once again appear this form of modelling has much to offer a full solution to this long-standing matter.

Over the past 7 years a series of fascinating results have been obtained from the high temperature superconducting (HTSC) cuprates using energy-resolved scanning tunnelling microscopy (SI-STM or STS) [1-7]. The latest in this series is published in the Feb 13th issue of Science by Hanaguri and coworkers [8]. I wish here to provide a different interpretation of their results than that which they pursue – a development of the one first introduced by Davis and coworkers [1,2], involving quasiparticle interference (QPI) and scattering by Bogoliubovons. My own interpretation falls within the general perspective afforded by the boson-fermion crossover mechanism for HTSC behaviour. This mechanism I have advocated for many years [9-18], but has been one remarkably overlooked by most workers in the field.

The scheme I have developed sees the phenomena recorded in the STS experiments as relating not to the Bogliubovons, but to quasiparticle scattering from a local-pair-derived mode appertaining to those bosons that reside outside the superconducting condensate. In the negative-$U$ driven resonant crossover approach, there occur both condensed and uncondensed bosons, besides residual uncondensed fermions. The bosonic pairs are perceived as a mixture of local pairs of symmetry $A_{1g}$, centred upon the axial antinodes, and induced BCS-like pairs of related $B_{1g}$ symmetry, predominant in the nodal or 45° directions [17]. The former through the negative-$U$ state binding energy, $U$, set the 'high energy' physics which comes to dictate the pseudogap state, whilst the induced pairs determine the low energy physics and the commonly ascribed $d$-wave gapping parameter, $2\Delta_{sc}$, for these HTSC systems once well away from the underdoped region [14,15]. The development of the two energy scales has been set out in close detail in [15] and [17], detail that is determined not only by the hole doping level, $p$, in these mixed-valent systems, but also by the differing levels of ionicity and screening dictated via the various 'counter-ions' in play [18]. The energies of the states and of the scattering processes between them are able to be tracked in detail not only via the STM experiments, but also in complementary fashion through the large bank of data available now from angle-resolved photoemission (ARPES) [19-23]. The interpretation offered below of the new STS results is felt to be a much more straightforward route to understanding than the one being developed in [1-8]. The latter takes a more conventional view of the superconductivity prevailing within HTSC systems than my own.

The energy-resolved STM experiments involve tunnelling into the sample being very finely scanned with the tip held at fixed energy (~ 1 to 50 meV) below $E_F$, and recording the local injection and extraction currents. Any real space signal component that is oscillatory is able subsequently to be Fourier analyzed through into **k**-space. The conductance ratio $Z(r,E)$ = $g(r,E)/g(r,-E)$ affords particularly appropriate real-space input in the case of Na-CCOC because it effectively suppresses the strong checkerboard component there within the overall scattering signal. The new work from Hanaguri *et al* [8] also introduces difference maps $s(r,E)$ to sharpen up changing features in the real space image under changing conditions.

This latest paper from Hanaguri *et al* [8] has pushed the previous probing of this fascinatingly detailed scattering a step further now by conducting the tunnelling in a strong *c*- axis magnetic field (of 5 or 11 tesla). Very significant changes are brought to the $k$-space scattering maps, certain features growing relatively in intensity, whilst others are weakened. The resolved features in $k$-space of especial interest here to the HTSC problem arise in octet groupings sited around ring loci. The geometry has been set out in figure 2 of ref. [17] (but beware vector labelling is different there). For any starting Fermi wavevector $k_F'$, the transfer vectors $q_1'$ to $q_7'$ to symmetry equivalent states - prior to a magnetic field being applied - mark the potential quasi-elastic scattering processes for the injected quasiparticles from the one such state to another. From all these scatterings there can emerge a total of 28 (i.e. ½x8x7) dominant standing waves. When $E = 0$ the $q$-space locus generated by all these key scatterings automatically outlines the basal geometry of the Fermi surface. Although as tunnelling energy -|$E$| is increased the individual dominant wavevectors $q_i$ are found to disperse, the Fermi surface geometry continues to be adhered to experimentally by the STS data, although note the scattering is quite diffuse. It has become the customary view [5-8] to associate the active bound scattering states here with the Bogoliubovons, the operating temperature being at 1.6 K very much less than $T_c$ (= 25 K for the slightly underdoped $p \sim x$ = 0.14 sample of Na-CCOC here employed). My own picture would instead see the states so strongly scattering the quasiparticles as being the excited mode states deriving from the local pair population, those pairs most stable at the antinodes. The uncondensed boson modes disperse from the antinodal points linearly upwards away from the local pair binding energy $U$ (in $p$=0.14 Na-CCOC ≈ 65 meV) while adhering to the parent quasi-particle $k$-space geometry of the Fermi surface [24]. Reflecting the $U$ values operative the excited bosonic modes become decreasingly bound, as the doping level (i.e. screening) in the examined material is pushed up and correspondingly as the ionicity level introduced via the particular counter-ions is diminished. The overall pattern of modal change as it is disclosed by SI-STM, ARPES and neutron scattering observations has been displayed in figures 3 and 4 of [17].

A recent STS study from Kohsaka, Davis and colleagues [7] of the $p$-dependence to the BSCCO-2212 $k$-space patterning affirms that the modal gradient grows as $p$ is reduced. Simultaneously the extent around the near-circular F.S. about the B.Z. corner over which STS data are forthcoming is observed to contract. The STS signal always is lost as the mode rises through $E_F$, and for underdoped material that occurs well in advance of the nodal 45° directions. The STS signal vanishes too around the antinodal/'hot spot' locations, where the scattering physics has become highly incoherent (for details in evidence from high-field transport data, see [18]). Note the above dispersion-line segments in fact recorded in the STS experiments are effectively linear and they do *not* bend around in Bogoliubovon fashion towards the nodal point as energy $E \rightarrow 0$, as has so often been claimed. The STS data now published in [8] by Hanaguri *et al* have been obtained down to $E$ = 1 meV, and figures 2 and 3 there relate to just 4 meV.

Part B of figure 4 in [8] reveals that the action of a magnetic field is to steepen the mode implicated and to drive the intercept with $E_F$ even further from the nodal direction. This is readily understood within boson-fermion crossover modelling. The effect of the field is to disturb the equilibrium between the pair and single quasi-particle population in favour of the latter. That is pretty standard and is directly evident in the rise in conductance $g(r,E=0)$ secured by application of the field (see figs 4C and D). The mode becomes steeper because the liberated quasi-particles here are closer to Mott insulation. The effect of a magnetic field of this magnitude on underdoped HTSC material has earlier been seen to advance the fluctuating segregation into 2D stripe domains and to enhance the degree of carrier localization within the $d^9$ domains [25,26,18]. The action of the field becomes accordingly to reduce the angular range over which the mode is registered, just as with the samples of decreasing $p$ reported upon previously by Kohsaka et al in [7]. [Note that in the latter work the various plots of the mode have been subject to different 'zero offsets': N.B. I would prefer to see all of these modal dispersion plots inverted, top-to-bottom, as in fig. 2 of [12] and figs. 3 and 4 of [17].) This steepening of the mode mirrors the effects from increased localization and scattering incoherence, much in evidence in the pseudogap state above $T_c$ within the ARPES results of Kanigel et al [20]. Below $T_c(H=0)$ comparable fascinating changes to quasiparticle/quasiparticle and quasiparticle/boson scattering have been met with recently by Hussey and colleagues [26] in high-field transport work. The latter I have interpreted at length in [18]. Within the underdoped regime the intensity of $e$-$b$ scattering is seen to be falling off rather sharply with decreasing $p$. This is because the boson population is not being sustained due to the chronic scattering (both $e$-on-$e$ and $e$-$b$ in nature) within the now more ionic material and the encroachment of incoherence in the quasiparticle system as one draws towards the Mott-Ioffe-Regel limit. A decreasing population of *induced* BCS nodal bosons occurs as the *local-pair* bosons relax in energy down away from the crucial resonant boson-fermion Feshbach condition. At $x = 0.14$, however, that resonance has not yet been so far departed from that $e$-$b$ scattering is not still a very significant process, as the current STS observations make clear.

This leaves us now with perhaps the most striking and significant observation forthcoming from the new research by Hanaguri et al [8], the intensity changes effected upon the STS standing waves by the application of magnetic fields of the above stated magnitudes. Spots $q_i = q_1, q_4, q_5$ grow relatively in intensity, whilst $q_i = q_2, q_3, q_6, q_7$ fall, although observe that the positions of all the spots remain relatively unchanged. (Beware here the rotation by $45^o$ between Hanaguri's figures 1 and 2). Therefore the changes are to the standing wave amplitudes themselves and the charge flows involved in their establishment. In the absence of an applied field, the approximate equality in intensity of all the observed spotting, whether from standing waves that run near-axially like $q_1$, $q_4$, $q_5$, or those that are near-nodal like $q_2$, $q_3$, $q_6$, $q_7$, manifests that the strong $e$-$b$ scattering is itself fairly isotropic, as befits the $A_{1g}$ symmetry of the local pair states. Hanaguri et al [8], because they are looking towards BCS Bogoliubovons of $B_{1g}$ symmetry to supply the primary signal, suggest that this experimental isotropy must be the consequence of impurities and gap inhomogeneities (known to be endemic in HTSC material, largely from STM work itself). It is in this cause of seeking to acquire some direct external control over the scattering that Hanaguri et al have turned to the effect of a magnetic field and specifically to the presence and action there of vortices within the superconducting condition above $H_{c1}$.

In HTSC material, especially near optimal doping, the coherence length $\xi$ is only ~20 Å or $5a_o$, and accordingly the entire effective diameter of a $\phi_o$ = $h/2e$ flux vortex (being ~ $5\xi$) will be about 100 Å. At a field of 11 tesla the spacing between vortices is evaluated to be just about 200 Å, and hence about a quarter of the field of view in the STS experiment should become taken up by the vortices, as may be seen in the $s(r,E,H)$ image of fig. 2c provided in Hanaguri's paper. The vortices in these circumstances ought to have then a significant impact on the amplitudes attained by the injected current standing-waves, with possible contributions both from the gap inhomogeneity level which the vortices sustain and the Doppler effect that they create in the charge superflow velocities around the cores. Note, nonetheless, the key scattering conditions within and outside the vortices must remain sufficiently similar for the standing wave wavelengths to remain essentially unaltered.

Now in the crossover modelling the low energy symmetry and charge flow are dominated by the nodal superconductivity of the *induced* BCS pairing, $B_{1g}$ in form. Anisotropy is able through this means to become introduced into the problem between the 45° (nodal) and the axial (antinodal) tunnelling initiated activity. This is what, it is felt, is being recorded in the STS field experiments, where, if the conductance signal Fourier-analysis is performed only over the sub-space allocated to the vortices the near-nodal sub-set of spots (i.e. 2,3,6,7) is found to be suppressed in relation to the near-axial sub-set (i.e. 1,4,5), whilst, if repeated over the sub-space allocated to the 'matrix' region *between* vortices, the reverse holds true. Inside the votices scattering off the axially stable local pairs is seemingly promoted. The authors of [8] would however look to superconducting coherence effects in the primary scattering, as well as, it would appear, the secondary to account for these observations. They are forced into this position by only entertaining a simple uniform BCS-like superconductivity with $B_{1g}$ *d*-wave ($x^2-y^2$) symmetry. Although at first glance that would seem the simpler position it in fact proves to engender unnecessary and inappropriate complication, as here when inserting the effects of a magnetic field into the FT-STS story. Specifically recall that in both NMR and optical experiments on HTSC materials there is in fact a marked *absence* of 'standard' coherence effects. Indeed the observation of such effects is not really to be expected with non-*s* wave superconductors given their momentum-dependent gapping.

The present STS work accordingly provides yet a further example of a quite sophisticated and discriminatory experimental situation in which the data are better engaged with by the negative-*U*, boson-fermion crossover account of events than by a more standard BCS-looking description. A comparable conclusion repeatedly has been reached, when for example discussing Gedik *et al*'s laser pump/probe crystallography (27,16), Kaminski *et al*'s magnetic circular dichroism results [28,15], Corson *et al*'s time domain spectroscopy results [29,11], or the ARPES of refs.[19] and [20], amongst many others [14,15].

Thanks are due to the University of Bristol for their continued support.